\documentclass[useAMS,usenatbib,usegraphicx,fleqn]{mn2e}

\usepackage{times} %

\usepackage{natbib}
\usepackage{amssymb,amsmath}
\usepackage{color}
\usepackage[colorlinks=true]{hyperref}
\usepackage{epstopdf} 
\usepackage{graphicx}
\usepackage[plain]{fancyref}
\usepackage[nolist]{acronym}

\makeatletter
\newcommand{\extraclearlabels}{\protected@write\@auxout{}{%
  \string\reset@newl@bel
}}
\makeatother

\graphicspath{{./}}

\newcommand{\planck}{{\it Planck}}

\newcommand{\rfive}{\theta_{500}}
\newcommand{\msun}{\mathrm{M}_\odot}

\newcommand{\lstar}{L^{\star}}

\newcommand{\griz}{\textsl{griz}}
\newcommand{\ysz}{Y_\text{SZ}}
\newcommand{\gt}{>}
\newcommand{\lt}{<}
\newcommand{\mr}[1]{\mathrm{#1}}
\newcommand{\zlimit}{$z_{\text{lim}(10^{15})}$}

\newcommand{\Munich}{$^{1}$}
\newcommand{\ExcellenceCluster}{$^{2}$}
\newcommand{\MPE}{$^{3}$}
\newcommand{\UoH}{$^{4}$}
\newcommand{\Durham}{$^{5}$}

\newcommand{\Princeton}{$^{7}$}

\newcommand{\Harvard}{$^{6}$}
\begin{acronym}
\acrodef{panstarrs}[Pan-STARRS]{Panoramic Survey Telescope \& Rapid
  Response System}
\acrodef{sze}[SZE]{Sunyaev-Zel'dovich effect}
\acrodef{cmb}[CMB]{cosmic microwave background}
\acrodef{ps1}[PS1]{Pan-STARRS1}
\acrodef{snr}[SNR]{Signal-to-Noise Ratio}
\acrodef{photoz}[photo-$z$]{photometric redshift}
\acrodef{rs}[RS]{Red Sequence}
\acrodef{rms}[RMS]{root mean square}
\acrodef{spt}[SPT]{South Pole Telescope}
\acrodef{act}[ACT]{Atacama Cosmology Telescope}
\acrodef{sdss}[SDSS]{Sloan Digital Sky Survey}
\acrodef{bcs}[BCS]{Blanco Cosmology Survey}
\acrodef{rcs}[RCS]{Red-Sequence Cluster Survey}
\acrodef{hon}[HON]{Halo Occupation Number}
\acrodef{bcg}[BCG]{Brightest Cluster Galaxy}
\acrodef{rass}[RASS]{{\it ROSAT} All Sky Survey}
\acrodef{wise}[WISE]{Wide-field Infrared Survey Explorer}
\acrodef{hod}[HOD]{Halo Occupation Distribution}
\acrodef{psf}[PSF]{Point Spread Function}
\acrodef{des}[DES]{Dark Energy Survey}
\acrodef{fwhm}[FWHM]{Full Width at Half Maximum}

\end{acronym}

\title[Optical Follow-up of Planck Cluster Candidates with PS1]{Optical Confirmation and Redshift Estimation of the Planck Cluster Candidates overlapping the Pan-STARRS Survey}
\author[J. Liu, et al.]{J.~Liu\Munich$^,$\ExcellenceCluster,
C.~Hennig\Munich$^,$\ExcellenceCluster,
S.~Desai\Munich$^,$\ExcellenceCluster,
B.~Hoyle\Munich,
J.~Koppenhoefer\MPE$^,$\Munich,
J.~J.~Mohr\Munich$^,$\ExcellenceCluster$^,$\MPE,
K.~Paech\Munich,
\newauthor 
W.~S.~Burgett\UoH,
K.~C.~Chambers\UoH,
S.~Cole\Durham,
P.~W.~Draper\Durham,
N.~Kaiser\UoH,
N.~Metcalfe\Durham,
\newauthor
J.~S.~Morgan\UoH,
P.~A.~Price\Princeton,
C.~W.~Stubbs\Harvard,
J.~L.~Tonry\UoH,
R.~J.~Wainscoat\UoH,
C.~Waters\UoH
\\
\Munich Department of Physics, Ludwig-Maximilians-Universit\"{a}t, Scheinerstr.\ 1, 81679 M\"{u}nchen, Germany \\
\ExcellenceCluster Excellence Cluster Universe, Boltzmannstr.\ 2, 85748 Garching, Germany \\
\MPE Max-Planck-Institut f\"{u}r extraterrestrische Physik, Giessenbachstr.\ 85748 Garching, Germany \\
\UoH Institute for Astronomy, University of Hawaii at Manoa, Honolulu, HI 96822, USA\\
\Durham Department of Physics, Durham University, South Road, Durham DH1 3LE, UK\\
\Harvard Department of Physics, Harvard University, Cambridge, MA 02138, USA\\
\Princeton Department of Astrophysical Sciences, Princeton University, Princeton, NJ 08544, USA\\
}

\begin{document}

\pdfpageheight 11.7in
\pdfpagewidth 8.3in

\maketitle

\begin{abstract}
\extraclearlabels   
We report results of a study of \planck\ \ac{sze} selected galaxy
cluster candidates using the \ac{panstarrs} imaging data.  We first
examine 150 \planck\ confirmed galaxy clusters with spectroscopic
redshifts to test our algorithm for identifying optical counterparts
and measuring their redshifts;  our redshifts have a typical accuracy
of $\sigma_{z/(1+z)} \sim 0.022$ for this sample.  Using 60 random sky locations,  we estimate that our chance of contamination through a random superposition is $\sim$ 3 per cent.  We then examine an additional 237 \planck\ galaxy cluster candidates that have no redshift in the source catalogue. Of these 237 unconfirmed cluster candidates we are able to confirm 60 galaxy clusters and measure their redshifts.  A further 83 candidates are so heavily contaminated by stars due to their location near the Galactic plane that we do not attempt to identify counterparts.  For the remaining 94 candidates we find no optical counterpart but use the depth of the \acl{ps1} data to estimate a redshift lower limit \zlimit\ beyond which we would not have expected to detect enough galaxies for confirmation.
Scaling from the already published \planck\ sample, we expect that $\sim$12 of these unconfirmed candidates may be real clusters. 

\end{abstract}


\acresetall
\section{Introduction}\label{sec:ps-introduction}

Massive clusters of galaxies sample the peaks in the dark matter density field, and analyses of their existence, abundance and distribution enable constraints on cosmological parameters and models \citep[e.g.][]{white93b,eke96,vikhlinin09,mantz10,rozo10,williamson11,2012JCAP...02..009H,2013MNRAS.434..684M,bocquet14}.  Surveys at mm wavelengths allow one to discover galaxy clusters through their \ac{sze}, which is due to inverse Compton interactions of \ac{cmb} photons with the hot intracluster plasma \citep{sunyaev70,  sunyaev72}.   Since the first \ac{sze}-discovered galaxy clusters were reported by the South Pole Telescope (SPT) collaboration \citep{staniszewski09}, large solid angle surveys have been completed, delivering many new galaxy clusters \citep[][]{reichardt13, planck13-29, hasselfield13}.  

The \ac{sze} observations  alone do not  enable one to determine the cluster redshift, and so additional followup data are needed.  In previous X-ray surveys, it was deemed necessary to obtain initial imaging followed by measurements of spectroscopic redshifts for each cluster candidate \citep[e.g.][]{rosati98,boehringer04,2012MNRAS.423.1024M}.  In ongoing\ \ac{sze} surveys, the efforts focus more on dedicated optical imaging \cite[e.g.][]{song12b, planck13-29} to identify the optical counterpart and measure photometric redshifts.  In the best case one leverages existing public wide field optical surveys such as the \acl{sdss} \citep[][]{york00}, the \acl{rcs} \citep{gladders05} or the \acl{bcs} \cite[][]{desai12}. \acused{sdss} \acused{rcs} \acused{bcd}

\begin{figure*}
\includegraphics[width=6.5in]{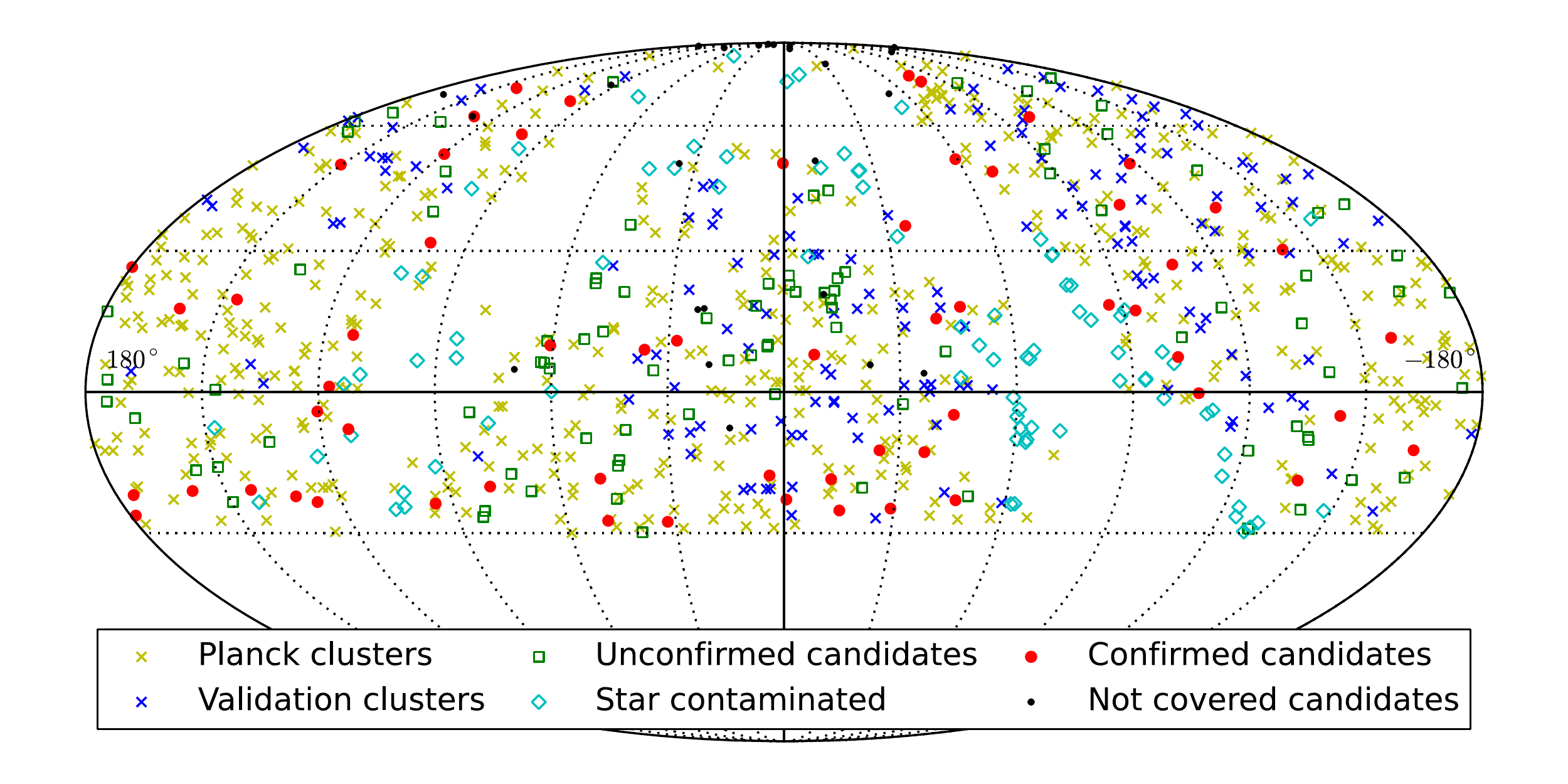}
\caption[The sky distribution of \planck\ clusters and candidates within the PS1 region.]{The sky distribution of \planck\ clusters and candidates within the \acs{ps1} region.  The crosses are previously confirmed \planck\ clusters, and the blue crosses mark the validation sample we use in this analysis.  For the remainder of the sample of previously unconfirmed \planck\ candidates, black dots mark those that are not fully covered by \acs{ps1} data,  red circles are clusters we confirm (see \Fref{tab:ps-confirmed}),  cyan diamonds are candidates that lie in areas of heavy star contamination, and green squares are candidates we do not confirm (see \Fref{tab:ps-unconfirmed}.)}
\label{fig:ps-whole-sky}
\end{figure*}

In March 2013 the \planck\ Collaboration released an \ac{sze} source catalogue with 1227 galaxy cluster candidates from the first 15 months of survey data \cite[]{planck13-29}.  Given the full-sky coverage of the \planck\ satellite, there is no single survey available to provide confirmation and redshift estimation for the full candidate list.  Of this full sample, 683 \ac{sze} sources are associated with previously known
clusters (e.g. Meta-Catalogue of X-ray detected Clusters of galaxies, \citealt{piffaretti11}; MaxBCG catalogue, \citealt{koester07a}; GMBCG catalogue, \citealt{hao10}; AMF catalogue, \citealt{szabo11}; WHL12 catalogue ,\citealt{wen12}; and SZ catalogues from \citealt{williamson11,reichardt13,hasselfield13}) and 178 are confirmed as new clusters, mostly through targeted follow-up observations.  The remaining 366 \ac{sze} sources are classified into three groups depending on the probability of their being a real galaxy cluster.  

In this paper we employ proprietary\ \ac{panstarrs} imaging data and a blinded analysis~\citep{Roodman05} to perform optical cluster identification and to measure photometric redshifts of \planck\ cluster candidates.  For those candidates where no optical counterpart is identified,  we provide redshift lower limits that reflect the limited depth of the optical imaging data.

This paper is organised as follows: we briefly describe the \ac{sze} source catalogue in \Fref{sec:ps-data-planck} and the optical \ac{panstarrs} data processing in \Fref{sec:ps-data-panstarrs}.  In \Fref{sec:ps-method} we provide the details of the \acl{photoz} (\acs{photoz} hereafter\acused{photoz}) estimation and cluster confirmation pipeline. Results of the \ac{photoz} performance and the confirmation of \planck\ candidates are presented in \Fref{sec:ps-result}.

\section{Data Description}\label{sec:ps-data-description}
We briefly describe the \planck\ \ac{sze} source catalogue in \Fref{sec:ps-data-planck} and refer the reader to the cited papers for more details.  In \Fref{sec:ps-data-panstarrs} we then describe the \ac{panstarrs}  optical data and calibration process we use to provide the images and calibrated catalogues needed for the cluster candidate follow up.

\subsection{Planck SZE Source Catalogue}\label{sec:ps-data-planck}

The \planck\ \ac{sze} source catalogue contains 366 unconfirmed cluster candidates, and it is available for download\footnote{\url{http://pla.esac.esa.int/pla/aio/planckProducts.html}.}.  This catalogue is described in detail elsewhere \cite[see][]{planck13-29}.  In summary, the \planck\ \ac{sze} sources are the union of detections from three independent pipelines, which are compared extensively in \cite{melin12}. The pipelines, which are optimized to extract the cluster \ac{sze} signal from the \planck\ \ac{cmb} data, are drawn from two classes of algorithms, namely two Matched-Multi filter pipelines, which are multi-frequency matched filter approaches \cite[]{melin06}, and the PowellSnakes pipeline, which is a fast Bayesian multi-frequency detection algorithm \cite[]{carvalho12}.  

The `union sample' is the combination of detections from each of these three pipelines with a measured \ac{snr} above $4.5$. Detections are further merged if they are within an angular separation of  $\leq5$~arcmin.  The detection, merging and combination pipelines have been tested using simulations and achieve a purity of 83.7 per cent \cite[]{planck13-29}.  With a sample of 1227 cluster candidates, we estimate that approximately 200 ($\sim 1227\times (1-83.7\%)$) are noise fluctuations. 
Thus, we expect to find 200 false detections in the remaining 366 unconfirmed candidates, indicating that the probability of a unconfirmed candidate to be a bona-fide cluster is only $(1-200/366)\sim45$~per cent.

The candidates in the union sample are grouped into three classification levels according to the likelihood of being a cluster. Class~1 is for high-reliability candidates that have a good detection in the \ac{sze} and are also associated with \acl{rass} \cite[\acs{rass};][]{voges99} \acused{rass} and \acl{wise} \cite[\acs{wise};][]{wright10} \acused{wise} detections.  The Class~2 candidates meet at least one of the three criteria in Class~1.  The Class~3 candidates correspond to low-reliability candidates that have poor \ac{sze} detections and no clear association with \ac{rass} or \ac{wise} detections.  A total of 237 unconfirmed \planck\ cluster candidates (Class~1, 2 and 3) lie within the \ac{panstarrs} footprint with enough coverage (c.f. \Fref{fig:ps-whole-sky} and \Fref{sec:ps-data-retrieval}).

The union sample also contains redshifts for previously known and confirmed clusters.  We create a validation sample by randomly selecting 150 of these clusters that fall within the \ac{panstarrs} footprint and have quoted \planck\ redshift uncertainties of $\lt0.001$.  We combine these 150 confirmed clusters with the sample of 237 cluster candidates for a total sample of 387 clusters and candidates.  We subject all targets in our total sample to the same procedure.  This blind analysis of our optical confirmation and \ac{photoz} estimation pipelines enables an important test of our methods as well as the characterisation of our photometric redshift uncertainties.  Note that the heterogeneous nature of \planck\ confirmation may result in a different redshift and mass distribution of the validation sample from that of unconfirmed clusters, but we do not expect this to lead to any important bias.  In what follows we refer to both confirmed clusters and cluster candidates within this total combined sample as `candidates'. 

\begin{figure}
\includegraphics[width=3.3in]{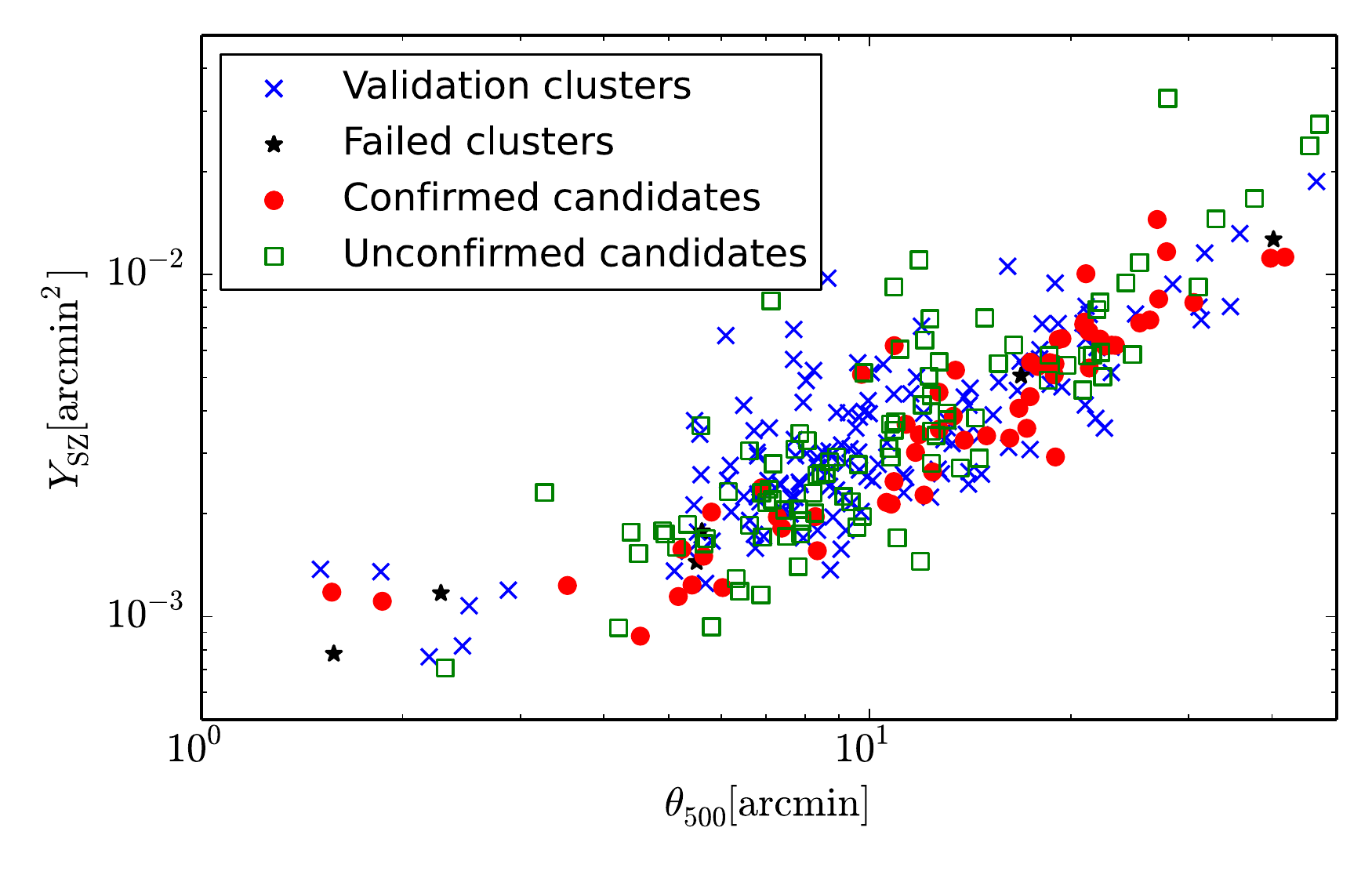}
\caption[The $\ysz--\rfive$ distribution of \planck\ clusters and candidates.]{The $\ysz\text{--}\rfive$ distribution of \planck\ clusters and candidates in our sample.  The \planck\ confirmed clusters are shown with blue crosses, and the six cases where our pipeline failed to confirm the systems are marked with black stars (see \Fref{sec:ps-result} for more details).  The \planck\ candidates with \ac{ps1} data are shown with red circles if we are able to measure a corresponding photometric redshift and with green squares if not.}
 \label{fig:ps-ysz-r500}
\end{figure}

For each candidate we use the following additional information given by each of the three individual \ac{sze} detection pipelines: the candidate position (Right Ascension $\alpha$, Declination $\delta$), the position uncertainty, the best-estimated angular size ($\theta_\text{s}$), and the integrated \ac{sze} signal $\ysz$ from the $\theta_\text{s}\text{--}\ysz$ likelihood plane provided with the \planck\ data products.  Furthermore, we convert the size to an angular estimate of $\rfive= c_{500}\theta_\text{s}$, where the concentration is set to $c_{500}=1.177$ as used in the cluster detection pipelines \cite[]{planck13-29}.  This angular radius $\rfive$ corresponds to the projected physical $R_{500}$ within which the density is 500 times the critical density at the redshift of the cluster.  In \Fref{fig:ps-ysz-r500} we show the $\ysz-\rfive$ distribution of the combined sample used in this work.

\subsection{PAN-STARRS1 Data}
\label{sec:ps-data-panstarrs}

For each candidate we retrieve the single epoch detrended images from the \ac{ps1} data server %
 and use those data to build deeper coadd images in each band.  This involves cataloguing the single epoch images, determining a relative calibration, combining them into coadd images, cataloguing the coadds and then determining an absolute calibration for the final multi-band catalogues.  We describe these steps further below.

\subsubsection{Data Retrieval}\label{sec:ps-data-retrieval}
The \ac{panstarrs} \cite[]{kaiser02} data used in this work are obtained from a wide field 1.8 metre telescope situated on Haleakala, Maui in Hawaii.  The \ac{ps1} telescope is equipped with a  1.4 gigapixel CCD covering a 7 deg$^2$ field of view, and it is being used in the \ac{ps1} survey to image the sky north of $\delta=-30^\circ$.  The $3\pi$ survey is so named because it covers 75 per cent of the celestial sphere.  The \ac{ps1} photometric system is similar to the \ac{sdss} filter system with \textsl{g}$_\mr{P1}$, $\textsl{r}_\mr{P1}$, $\textsl{i}_\mr{P1}$, $\textsl{z}_\mr{P1}$,  $\textsl{y}_\mr{P1}$ (where \ac{sdss} had \textsl{u}), and a wide band $\textsl{w}_\mr{P1}$ for use in the detection of Near Earth Objects \cite[]{tonry12}.  In this study we process data from the first four filters and denote them as \griz.  

We obtain single epoch, detrended, astrometrically calibrated and warped \ac{ps1} imaging data \citep{metcalfe13} using the \ac{ps1} data access image server.  We use 3PI.PV2 warps wherever available and 3PI.PV1 warps in the remaining area.  We select those images that overlap the sky location of each candidate, covering a square sky region that is $\sim$1$^\circ$ on a side.  The image size ensures that a sufficient area is available for background estimation.

\subsubsection{Single Epoch Relative Calibration}

The subsequent steps we follow to produce the science ready coadd images and photometrically calibrated catalogues are carried out using the Cosmology Data Management system (CosmoDM), which has its roots in the Dark Energy Survey data management system \citep{ngeow06,mohr08,mohr12} and employs several AstrOMatic codes that have been developed by Emmanuel Bertin (Institut d'Astrophysique de Paris).  

We build catalogues from the \ac{ps1} warped single epoch images using  {\sc SExtractor}~\citep{Bertin96}.  The first step is to produce a model of the \ac{psf} variations over each of the input single epoch images.  This requires an initial catalogue containing stellar cutouts that are then built, using {\sc PSFex}~\citep{Bertin11}, into a position dependent \ac{psf} model.  With this model we then recatalog each image using model fitting photometry with the goal of obtaining high quality instrumental stellar photometry over each input image.

For each band, relative photometric calibration is performed using these catalogues;  we compute the average magnitude differences of stars from all pairs of overlapping images and then determine the relative zeropoints using a least squares solution.  The stars are selected from the single epoch catalogues using the morphological classifier {\tt spread\_model} \citep[e.g. in particular $\lvert \text{\tt spread\_model} \rvert < 0.002$; see][]{desai12, bertin13}.  We use the \ac{psf} fitting magnitude {\tt mag\_psf} for this relative calibration.

\begin{figure}
\begin{center}
\includegraphics[width=3.5in]{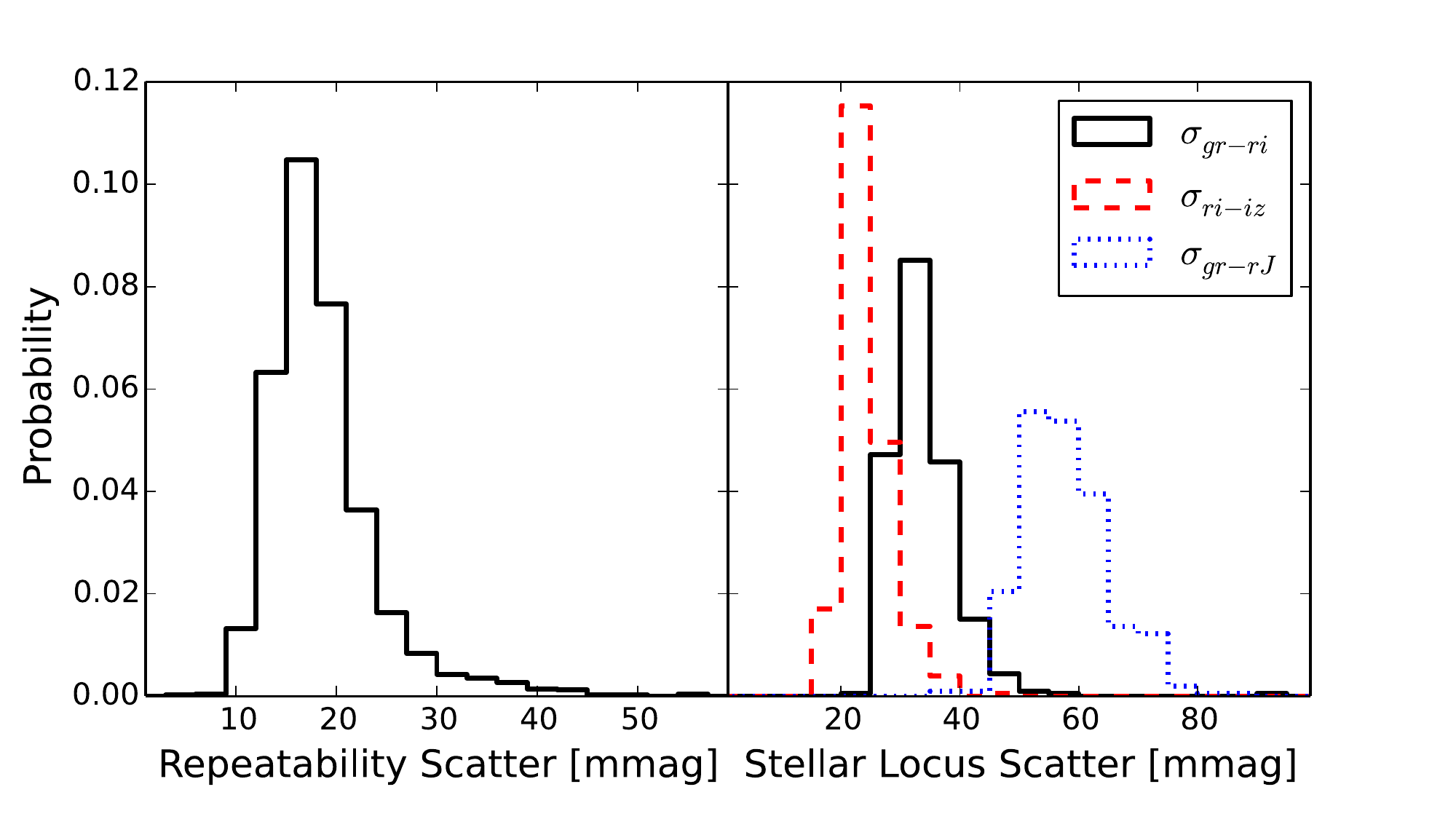}
\caption[Stellar locus scatter and repeatability of Pan-STARRS data.]{The left panel shows the histogram of single epoch repeatability scatter, extracted for bright stars in the full ensemble of candidates.  All bands have similar distributions, and so only the combined distribution is shown.  The median scatter is 16, 18, 19, and 17~mmag in \griz, respectively.  The right panel shows the histogram of the stellar locus scatter extracted from the full ensemble of 387 candidates.  The median values of the scatter distributions for all candidates are 34, 24, and 57~mmag in $\textsl{g--r}$ vs.\ $\textsl{r--i}$, $\textsl{r--i}$ vs.\ $\textsl{i--z}$ and $\textsl{g--r}$ vs.\ $\textsl{r--J}$ colour spaces.}
\label{fig:ps-qa}
\end{center}
\end{figure}

We test the accuracy of the single epoch model fitting relative photometry by examining the variance of multiple, independent measurements of stars.  \Fref{fig:ps-qa} contains a histogram of the so-called repeatability of the single epoch photometry.  These numbers correspond to the \ac{rms} variation of the photometry of bright stars scaled by $1/\sqrt{2}$, because this is a difference of two measurements.  We extract these measurements from the bright stars where the scatter is systematics dominated (i.e. the measurement uncertainties make a negligible contribution to the observed scatter).  We measure this independently for each band and candidate and use the behaviour of specific candidate tiles relative to the ensemble to identify cases where the single epoch photometry and calibration need additional attention.  The median single epoch repeatability scatter is 16, 18, 19, and 17 mmag in \griz, respectively.

As part of this process we obtain PSF \ac{fwhm} size measurements for all single epoch images.  The median \ac{fwhm} for the full ensemble of imaging over all cluster candidates is $1\farcs34$, $1\farcs20$, $1\farcs12$, and $1\farcs09$ in \griz, respectively.

\subsubsection{Coaddition, Cataloguing and Absolute Calibration}

\begin{figure} 
\includegraphics[width=3.3in]{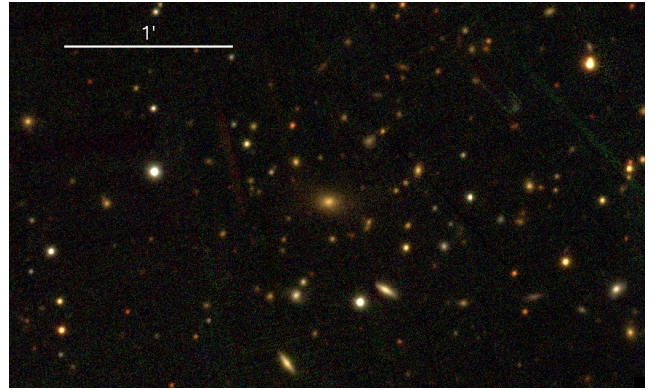}
\caption[Example pseudo-colour image in the {\textsl gri} bands of cluster candidate 218.]{Example pseudo-colour image in the {\textsl  gri} bands of cluster candidate 218.  In this case the \planck\ \ac{sze} candidate centre is about 4~arcmin away from the \ac{bcg}, which is at the centre of this image.  This exemplifies an extreme case of the large offset between the \planck\ centre and the \ac{bcg}.}
\label{fig:ps-pseudo-img}
\end{figure}

The coadd images are then generated from the single-epoch images and associated relative zero points.  For each candidate tile we generate both \ac{psf} homogenized and non-homogenized coadds. To create the homogenized coadds, we convolve the input warp images to a \ac{psf} described by a Moffat function with \ac{fwhm} set to equal the median value in the single epoch warps overlapping that candidate.  We homogenize separately for each band.  We then combine these homogenized and non-homogenized warps using {\sc SWarp}~\citep{bertin02} in a median combine mode.  We create a $\chi^2$ detection image~\citep{Szalay99} from the homogenized coadds using both %
 \textsl{i} and \textsl{z} bands. The \ac{psf} homogenized coadds are then catalogued using {\sc SExtractor} in dual image mode with this $\chi^2$ detection image.  We use {\sc SExtractor} in \ac{psf} correcting, model fitting mode.  The non-homogenized coadds are only used for visual inspection and for creating pseudo-colour images of the candidates (see \Fref{fig:ps-pseudo-img}).    For a more detailed discussion of coadd homogenization on a different survey dataset, see \citet{desai12}.

We use the stellar locus together with the absolute photometric calibration from the 2MASS survey \citep{skrutskie06} for the final, absolute photometric calibration for our data \citep[see also][and references therein]{desai12}.  For this process we adopt the \ac{ps1} stellar locus measured by \citet{tonry12}.  

In our approach we first apply extinction corrections to the relative photometry from the catalogues using the dust maps from \citet{schlegel98}.  This correction removes the overall Galactic extinction reddening, making the stellar locus more consistent as a function of position on the sky.  As is clear from \Fref{fig:ps-whole-sky}, the \planck\ cluster candidates extend to low galactic latitude, and some lie in locations of extinction as high as $A_V=1.8$~mags.  Most of the targets with $A_V>0.5$~mag also have very high stellar contamination, making it impossible for us to use the \ac{ps1} data for candidate confirmation.  \citet{high09} examined photometrically calibrated data lying in regions with a range of extinction reaching up to $A_V\sim1$~mag, showing that within this range the stellar locus inferred shifts are equivalent to the Galactic extinction reddening corrections to within an accuracy of $\sim$20~mmag.

We then determine the best-fit shifts in $\textsl{g--r}$ and $\textsl{r--i}$ that bring our observed stellar sample to coincide with the \ac{ps1} locus.  We repeat this procedure for $\textsl{i--z}$ while using the $\textsl{r--i}$ result from the previous step.  This allows for accurate colour calibration for the \ac{ps1} bands used for the cluster photometric redshifts. To obtain the absolute zeropoint, we adjust the $\textsl{g--r}$ vs. $\textsl{r--J}$ locus until it coincides with the \ac{ps1} locus.   This effectively transfers the $\sim$2~per cent 2MASS photometric calibration \citep{skrutskie06} to our \ac{ps1} catalogues.

\begin{figure}
\begin{center}
\includegraphics[width=3.5in]{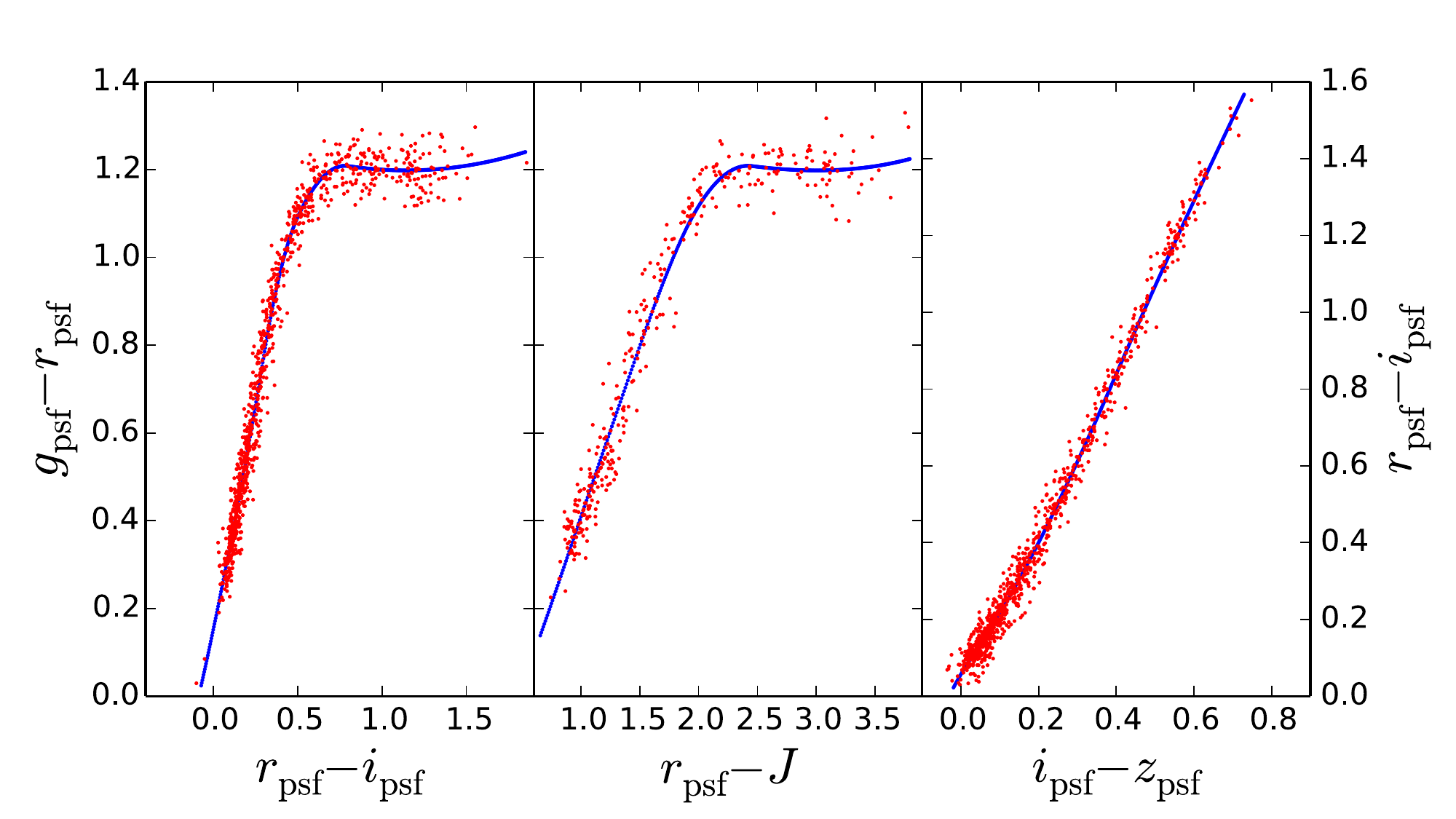}
\caption[The stellar loci in three different colour-colour spaces for the \planck\ cluster 307.]{The stellar loci in three different colour-colour spaces for the \planck\ cluster 307 are shown.  The blue line shows the \ac{ps1} stellar locus, and red points show \ac{psf} model fitting magnitudes of stars from our catalogues for this tile. We use the stellar locus for absolute photometric calibration.  The scatter about the stellar locus provides a good test of photometric quality; for this cluster the values of the scatter in $\textsl{g--r}$ vs.\ $\textsl{r--i}$ (left), $\textsl{g--r}$ vs.\ $\textsl{r--J}$ (middle) and $\textsl{r--i}$ vs.\ $\textsl{i--z}$ (right) colour spaces are 29, 48 and 17~mmag, respectively. }
\label{fig:slrfig}
\end{center}
\end{figure}

An illustrative plot of the stellar loci for \planck\ cluster 307 is shown in \Fref{fig:slrfig}.  The scatter of our model fitting photometry about the stellar locus provides a measure of the accuracy of the coadd model fitting photometry.  In the case of candidate 307 the scatter around the stellar locus in $\textsl{g--r}$ vs.\ $\textsl{r--i}$, $\textsl{g--r}$ vs.\ $\textsl{r--J}$, and $\textsl{r--i}$ vs.\ $\textsl{i--z}$ is 29, 48, and 17~mmag, respectively.  In \Fref{fig:ps-qa} we show the histogram of scatter for the ensemble of candidates in each of these colour--colour spaces. The median scatter of the stellar locus is 34, 24, and 57~mmag in $\textsl{g--r}$ vs.\ $\textsl{r--i}$, $\textsl{r--i}$ vs.\ $\textsl{i--z}$, and $\textsl{g--r}$ vs.\ $\textsl{r--J}$, respectively.  These compare favourably with the scatter obtained from the SDSS and BCS datasets \citep{desai12}.  Note that the shallow 2MASS photometry contributes significantly to the scatter in one colour-colour space, but in the others we restrict the stars to only those with photometric uncertainties $<$10~mmag (see \Fref{fig:ps-qa}).  We use the scatter measurements within each candidate tile together with the behaviour of the ensemble to identify any candidates that require additional attention.  We note that the PS1 ubercal calibration method \citep{schlafly12} has been able to achieve internal photometric precision of $<$ 10 mmag in photometric exposures in \textsl{g}, \textsl{r}, and \textsl{i} and $\simeq  10$ mmag in \textsl{z}, but it has not been applied over the whole 3PI dataset yet.

\begin{figure}
\includegraphics[width=3.4in]{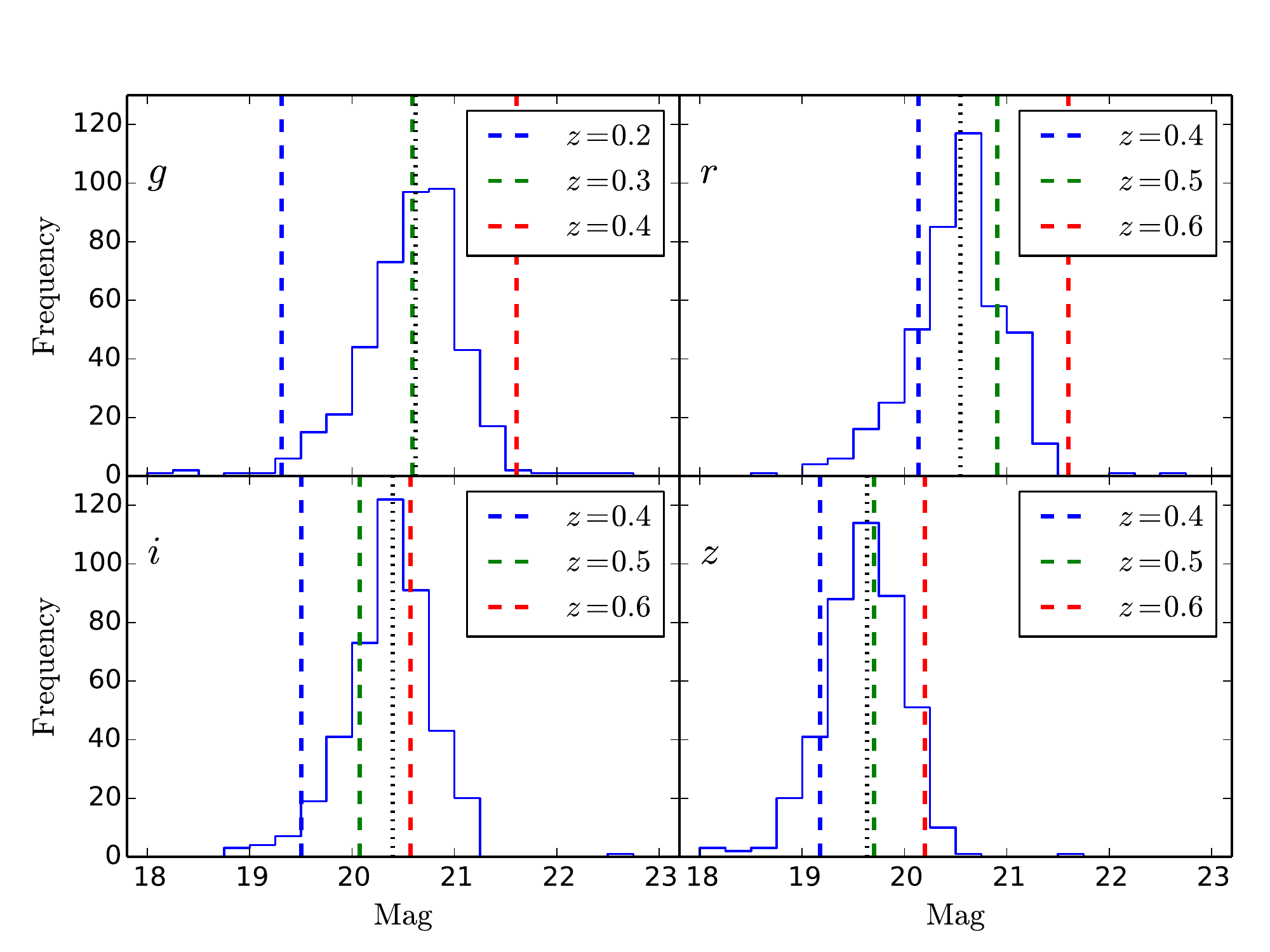}
\caption[The distributions of \griz\ band 10$\sigma$ depths ({\tt mag\_auto}) for PS1 fields around each \planck\ candidate.]{The distributions of \griz\ band 10$\sigma$ depths ({\tt mag\_auto}) for \ac{ps1} fields around each \planck\ candidate.  The dashed lines mark the magnitudes of $\lstar$ galaxies at different redshifts.  The dotted lines mark the median depths, which are 20.6, 20.5, 20.4 and 19.6 in \griz, respectively. The \ac{ps1} data are typically deep enough for estimating cluster redshifts out to or just beyond $z=0.5$ (see also \Fref{fig:ps-photoz-depth}).}
\label{fig:ps-depth}
\end{figure}

We estimate a photometric 10~$\sigma$ depth, above which the galaxy catalogue is nearly complete, in each coadd by calculating the mean magnitude of galaxies with {\tt mag\_auto} uncertainties of 0.1.  In \Fref{fig:ps-depth} we show the histograms of the distribution of depths in each band;  the median depths in \griz\ are 20.6, 20.5, 20.4 and 19.6 (denoted by dotted lines).  We note that the median depths are shallower than the limiting depths reported by the \ac{ps1} collaboration \cite[]{metcalfe13}, but this difference is mainly due to a different definition of the depth.  We find that to this depth the magnitude measurements from {\tt mag\_auto} and the colour measurements using {\tt det\_model} are well suited for the redshift estimation analysis which we describe in \Fref{sec:ps-field-depth-estim}.

Variation in observing conditions leads to non uniform sky coverage across the \ac{ps1} footprint.  One result is that the depth varies considerably from candidate to candidate; another is that not all candidates are fully covered in each of the bands of interest.  Overall 387 cluster candidates have been fully covered.  In \Fref{fig:ps-whole-sky} we show the sky distribution of our full sample together with that of the \planck\ sample.  
\section{Method}\label{sec:ps-method}
In this section we describe the optical confirmation and redshift estimation technique that we apply to the \ac{ps1} galaxy catalogues (see \Fref{sec:ps-redshifts-estimation}).  Then in \Fref{sec:ps-field-depth-estim} we describe the method we use -- especially in candidates without optical counterparts -- to estimate the redshift lower limit as a function of the field depth. 

\subsection{Confirmation and Redshift Estimation}
\label{sec:ps-redshifts-estimation} 
We employ the red sequence galaxy overdensity associated with a real cluster to 
identify an optical counterpart for the \planck\ candidates and to estimate a photometric redshift; our method follows closely that of \cite{song12a}, which has been applied within the \ac{spt} collaboration to confirm and measure redshifts for 224 \ac{sze} selected cluster candidates \citep{song12b} and then later for the full 2500~deg$^2$ SPT-SZ survey sample \citep{bleem14}.  A similar approach has been used to identify new clusters from optical multi-band surveys using only the overdensity of passive galaxies with similar colour \citep{gladders05}.  We start with additional information from the \ac{sze} or X-ray about the sky location and, in principle, also a mass observable such as the \ac{sze} or X-ray flux that can be used at each redshift probed to estimate the cluster mass and characterise the scale of the virial region within which the red sequence search is carried out (Hennig et al, in preparation). %
 We describe the procedure below.

We model the evolutionary change in colour of cluster member galaxies across cosmic time by using a composite stellar population model initialised with an exponentially decaying starburst starting at redshift $z=3$ with decay time $\tau=0.4$~Gyr \citep{bruzual03} .  We introduce tilt into the red sequence of the passive galaxies by adopting 6 models with different metallicities adjusted to follow the observed luminosity--metallicity relation in Coma \citep{poggianti01}.  Using the absolute \ac{ps1} filter transmission curves, which include atmospheric, telescope, and filter corrections \citep{tonry12}, as inputs for the package \textit{EzGal} \citep{mancone12}, we generate fiducial galaxy magnitudes in \griz\ bands over a range of redshifts and within the range of luminosities $3\lstar \geq L \geq 0.3\lstar$, where $\lstar$ is the characteristic luminosity in the \cite{schechter76} luminosity function.  

We exclude faint galaxies by employing a minimum magnitude cut of $0.3 \lstar$;
to reduce the number of junk objects in the catalogue we remove all objects with a magnitude uncertainty $\gt0.3$.  In \cite{song12b} a fixed aperture is used to both select cluster galaxies and perform background subtraction.  In this work we use the \planck\ derived radius $\rfive$ centred on the position of the candidate to separate galaxies into cluster and field components.  Galaxies located between$(1.5\text{--}3)\rfive$ are used to estimate background corrections.  Each galaxy within the radial aperture $\rfive$ is assigned two weighting factors. The first one is a Gaussian colour weighting corresponding to how consistent  the colours of the galaxy are with the modelled red sequence at that redshift. This red likelihood, $\mathcal{L}_\text{red}$, is calculated separately for each of the following colour combinations: $\textsl{g--r}$ and $\textsl{g--i}$, which are suitable for low redshift ($z<0.35$) estimation, and  $r\text{--}i$ and $\textsl{r--z}$, which are suitable for intermediate redshift ($0.35<z<0.7$) estimation. The second factor weights the galaxy depending on the radial distance to the cluster centre,  $\mathcal{L}_\text{pos}$, and for this function we adopt a projected NFW profile \cite[]{navarro97} with concentration $c=3$. In this way, all galaxies physically close to the cluster centre and with colours consistent with the red sequence at the redshift being probed are given higher weight. Conversely any galaxies in the cluster outskirts with colours inconsistent with the red sequence are given a small weight. 

The method then scans a redshift range $0\lt z\lt0.7$ with an interval $\delta z=0.01$ and iteratively recomputes the above weight factors using the modelled evolution of the red sequence.  For each cluster candidate we construct histograms of the weighted number of galaxies as a function of redshift for each above-mentioned colour combination.  The weighted number of galaxies is determined for each colour combination as the background subtracted sum of all galaxy weights at each given redshift.

For each cluster we identify the appropriate colour combination using a visual examination of the red sequence galaxies within the cluster centre and record the \ac{bcg} position, if possible.  The final \ac{photoz} is estimated by identifying the most significant peak in the background corrected likelihood histogram from all galaxies within $\rfive$.  The associated \ac{photoz} uncertainty is determined from the width of a Gaussian fit to the peak with outliers at $>3\sigma$ removed.   Specifically, the \ac{photoz} uncertainty $\delta z_\text{phot}$ is the standard deviation of the Gaussian divided by the square-root of the weighted galaxy number in the peak.  The performance is presented in the following section. We note that, given the depth of the data (see \Fref{fig:ps-depth}), we are unable to identify candidates with redshifts $z>0.7$.  

The optical confirmation and \ac{photoz} estimation break down if no significant peak is found in the likelihood histogram. In addition to the case where the candidate is not a cluster, there are three categories of failure that are possible: (1) those candidates with a \planck\ $\rfive$ that is so small such that there are not enough red sequence galaxies within the search aperture, (2) those that have a radius $\rfive$ above 30 arcminutes, in which case our standard $0.7^\circ\times0.7^\circ$ coadd catalogue region typically does not contain enough remaining area to measure the background well, and (3) those candidates that have a relatively large offset between the visually confirmed cluster centre and the \planck\ position.   Clusters with $\rfive>30''$ all lie at low redshift, where-- given the sensitivity of the \planck\ \ac{sze} selection -- we would expect these systems to have already have been confirmed by low redshift all sky surveys \citep[e.g.][]{abell58,abell89,voges99}.   For cases 1 and 2, we rerun the pipeline with a radius of 5~arcmin, which is the same as the \planck\ matching radius.  For the 3rd case we recenter at the coordinates of the \ac{bcg} if a \ac{bcg} can be identified within the coadd region. %
With the approach described above, the uncertainties associated with the \planck\ candidate position and size have no significant impact on our confirmation and \ac{photoz} estimation.  We demonstrate this with the validation sample in \Fref{sec:ps-method-validation}.

\begin{figure}
  \includegraphics[width=3.40in]{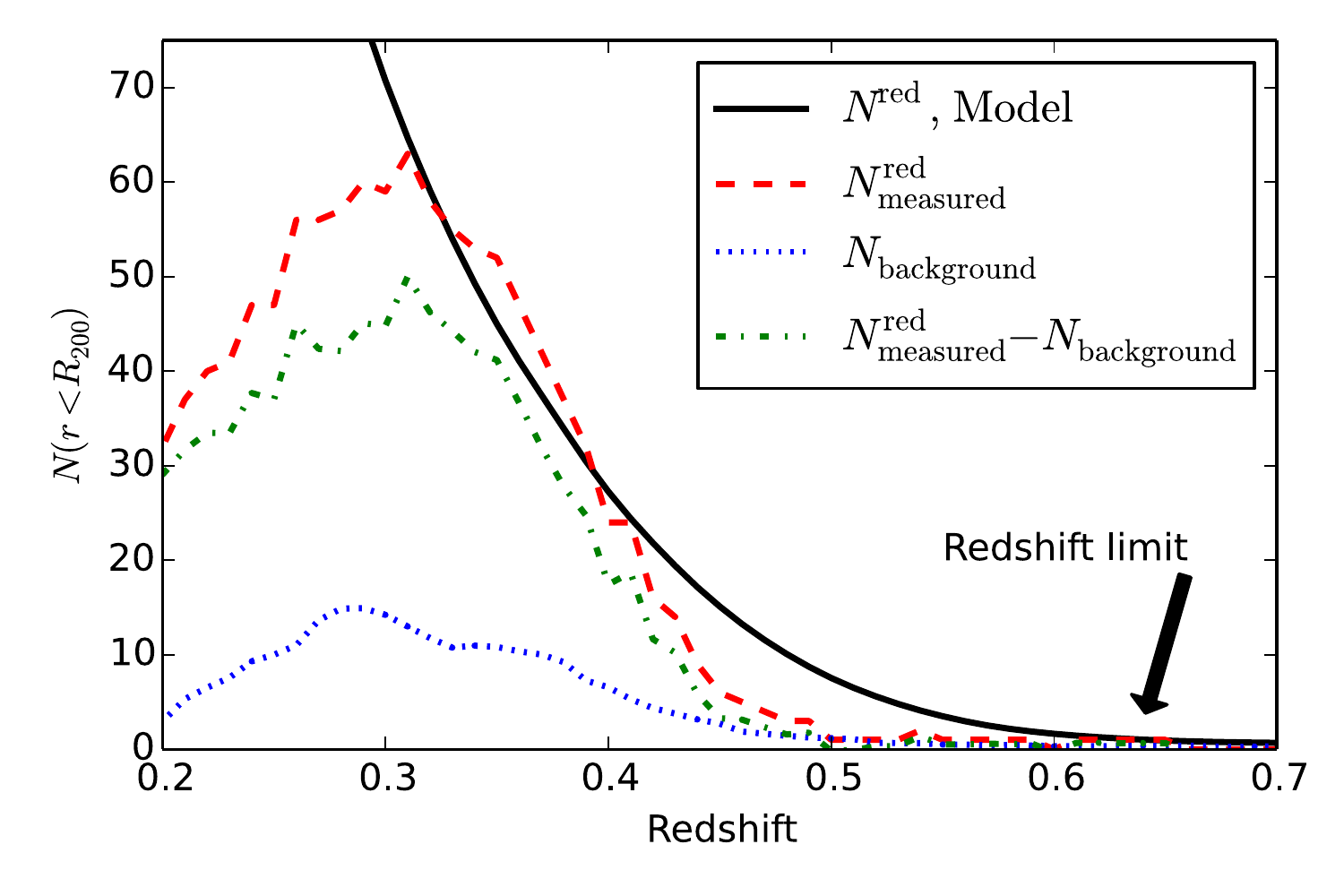}
  \caption[The observed number of red galaxies in the \planck\ confirmed cluster 442 at $z=0.3436$.]{The observed number of red galaxies in the \planck\ confirmed cluster 442 at $z=0.3436$. The red dashed line is the red sequence galaxy number within $R_{200}$; the blue dotted line is the background number corrected to the $R_{200}$ area of the cluster; and the green dash-dot line is the difference between those two.  The black line is the predicted number of red sequence galaxies $N^\mathrm{red}$, which increases towards lower redshift as more and more faint galaxies in the luminosity function slide above the imaging detection threshold.  We use this function together with the background to estimate a redshift lower limit in cases where no optical counterpart is identified.}
  \label{fig:ps-photoz-one-hon}
\end{figure}

\subsection[Redshift Lower Limits]{Redshift Lower Limits \zlimit}
\label{sec:ps-field-depth-estim}

For clusters where there is no obvious over-density of red sequence galaxies, there are two possibilities:  (1) the candidate is a noise fluctuation, or (2) the cluster is at high enough redshift that the \ac{ps1} imaging data is not deep enough to detect the cluster galaxy population.  Given the contamination estimates provided by the \planck\ collaboration, we expect approximately half of our candidates to be noise fluctuations.  However, of the 45 per cent that are real clusters we expect a small fraction of them to lie at redshifts too high to be followed up using the \ac{ps1} data.  In particular, the observed redshift distribution of the 813 previously confirmed \planck\ clusters has 3 per cent of those clusters lying at $z>0.60$, which is a reasonable expectation of the redshift limit to which we could expect to use \ac{ps1} data to confirm a cluster.  Simple scaling suggests we should expect approximately 3 clusters to lie at $z>0.6$ in our candidate sample.  Thus, for each of these undetected systems we calculate the minimum redshift \zlimit\ beyond which the candidate would be undetectable in our \ac{ps1} imaging.

To estimate the redshift lower limit we first measure the depth of the catalogue at the coordinates of the candidate (see \Fref{fig:ps-depth}) and then predict, as a function of redshift, the statistical significance of the detectable galaxy overdensity above background.  To do this we adopt a typical mass for a \planck\ cluster of $M_{200}=1\times10^{15}\msun$ and use a model for the \ac{hod} of red sequence galaxies in \ac{sze} selected clusters of this mass (Hennig et al, in preparation).  That analysis uses a joint dataset consisting of 74 SPT selected clusters and \ac{des} imaging of the galaxy populations for clusters with $M_{200}>4\times10^{14}\msun$ extending over the redshift range $0\lt z\lt 1.2$.  The results are in good agreement with those from a sample of $\sim100$ clusters studied in the local Universe \citep{lin04a}.

The estimated number of detectable red cluster galaxies $N_i^\text{red}(z)$ for candidate $i$ at redshift $z$ can be expressed as
\begin{equation}
  \label{eq:ps-hon}
  N_i^\text{red}(z) =  \Big[1+V\phi_{\star}(z)\int_{y_\text{L}}^{+\infty} y^{\alpha} e^{-y} \mr{d}y\Big]\times f_\mathrm{r}(z),
\end{equation}
where %
$\phi_{\star}(z)$ is the characteristic number density of galaxies, $\alpha$ is the faint end slope, $y=L/\lstar(z)$ where $\lstar(z)$ is taken from the passive evolution model used in this work, $V$ is the virial volume, and $y_\text{L}$ is the luminosity limit determined from the catalogue depth for the candidate.  For these parameters we adopt values that are consistent with the Hennig et al (in preparation) results.  Namely, we use  $\phi_{\star}(z)=3.6E(z)^{2} [\text{Mpc}^{-3}\text{mag}^{-1}]$ and 
$\alpha=-1.05(1+z)^{-2/3}$.  The number one comes from the fact that the \ac{bcg} is not included in this scaling relation, but needs to be counted in the \ac{hon}.  We additionally multiply by the red fraction, $f_\mathrm{r}(z)= 0.8 (1+z)^{-1/2}$, at the appropriate redshift.
Finally, we apply a correction to relate the number of galaxies within $R_{200}$ to the number of galaxies projected within $R_{500}$.  For this correction we adopt an NFW distribution of galaxies with concentration $c_{200}=3$.

The measured number of red galaxies is determined directly from the candidate catalogue as follows.  We set a magnitude error cut of 0.3 and a magnitude limit of $0.3 \lstar$ in analogy to the photo-z estimation and sum all galaxies with $\mathcal{L}_\text{red}>0.05$ projected within the $R_{500}$ radius, which is converted from the typical \planck\ mass cluster ($M_{200}=1\times10^{15}\msun$) using an NFW model with concentration $c$ \citep{duffy08}.  We set the centre of the candidate to be the visually identified \ac{bcg} position if it is available, or, alternatively, we use the \planck\ candidate centre.  The background number is extracted from the area beyond $3R_{500}$ and a correction for the differences in cluster search and background area is applied.

\begin{figure}
  \includegraphics[width=3.4in]{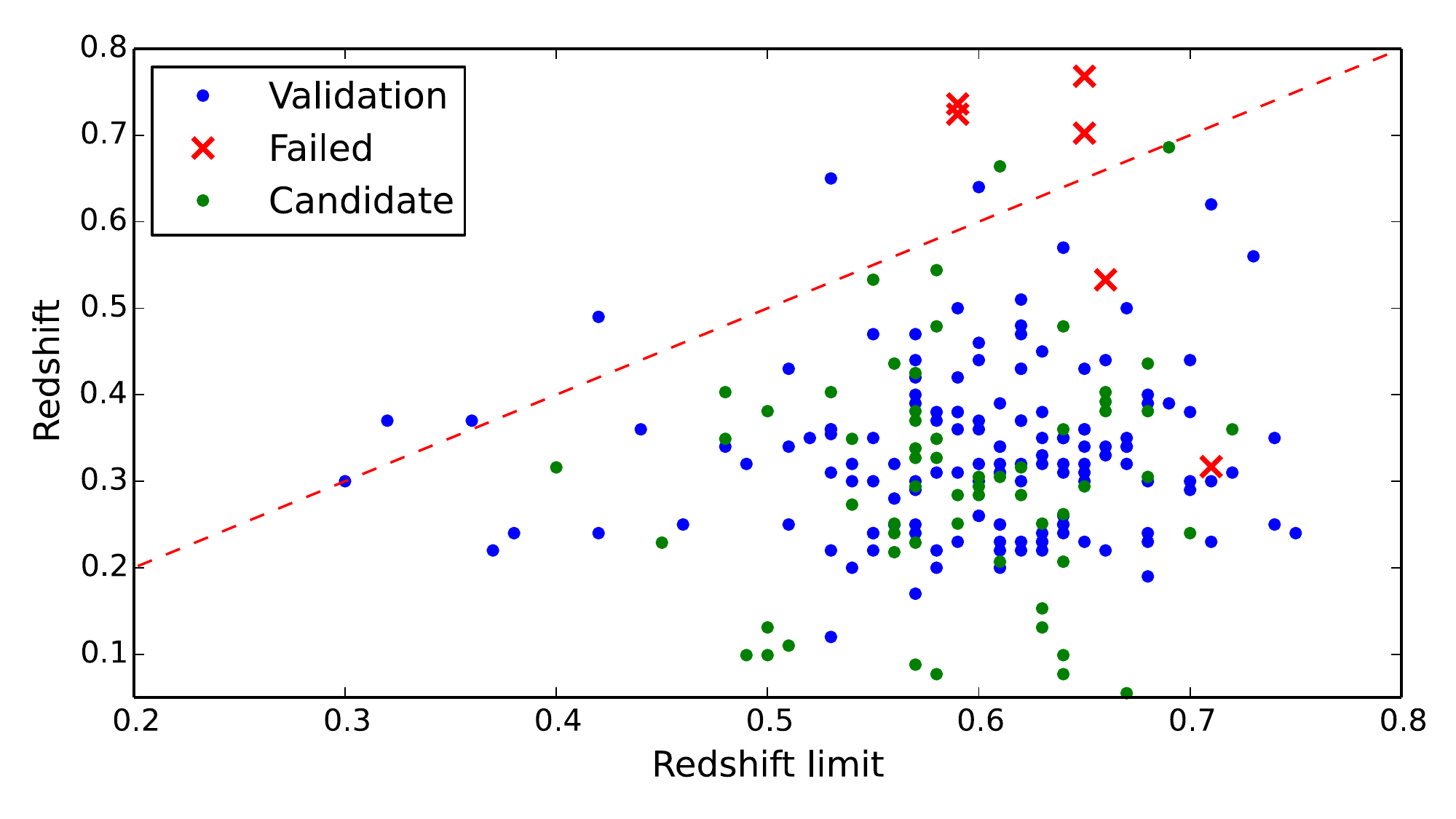}
  \caption[Redshift limits for the \planck\ sample.]{We plot the redshift lower limit \zlimit\ for a cluster with mass $M_{200}=1\times10^{15}\msun$ versus cluster photometric redshift for the clusters in the validation sample (blue points) and the clusters we have confirmed in \ac{ps1} (green points).  Six red crosses mark the systems in the validation sample (with spectroscopic redshifts) that we failed to confirm; we discuss these in \Fref{sec:ps-method-validation}.  Clusters below the red dashed line have the required \ac{ps1} imaging depth to enable a robust redshift measurement.  Those clusters above the line are marked as having shallow data in \Fref{fig:ps-photoz-valid}.}
  \label{fig:ps-photoz-depth}
\end{figure}

Given the individual catalogue depth, we estimate the redshift lower limit as the lowest redshift where the background galaxy population has at least a 5 per cent chance to be as large as that expected for a cluster of $M_{200}=1\times10^{15}\msun$.  That is, we require that the predicted cluster galaxy population be detectable above background at a minimum of 2\,$\sigma$.  We first calculate the \ac{hon} from \Fref{eq:ps-hon} for all redshifts (black line in \Fref{fig:ps-photoz-one-hon}); we then measure the number of red sequence galaxies in the background region and correct it for the difference in area between the cluster search and background region.  Finally, we find the highest redshift such that the cluster would be detected with 2\,$\sigma$ significance.  The depths for all candidates are plotted in \Fref{fig:ps-photoz-depth} and reported for each unconfirmed candidate in \Fref{tab:ps-unconfirmed}; the median redshift lower limit for our data is \zlimit$=0.60$.

\section{Results}\label{sec:ps-result}

We apply our method to the entire sample of 387 candidates in a uniform manner.   Thereafter, we examine the subset of candidates that are previously confirmed clusters to validate our method.  Our approach of blinding the sample eliminates any possible confirmation bias and allows us to accurately estimate the failure rate and to test our photometric redshift uncertainties.  In addition, we apply the same confirmation procedure over random sky regions to measure the probability of random superposition.  We then discuss the remaining candidates, presenting new photometric redshifts where possible.

\begin{table}
{\centering
\caption{Photo-$z$ comparison for \protect\cite{rozo14} sample.}\label{tab:ps-cmp-rozo}%
    \begin{tabular}{rrrrr}
    \hline\hline
  ID & \planck & SDSS & \ac{ps1} & Rozo's comment\\ \hline
    13    & 0.429 & \textbf{0.325} & 0.35  &  \\
    97    & 0.361 & \textbf{0.310} & 0.29  &  \\
    216   & \textbf{0.336} & \textbf{0.359} & 0.30  & Mismatch \\
    443   & 0.437 & \textbf{0.221} & 0.22  &  \\
    484   & 0.317 &   -  &    -   &  Unconvincing\\
    500   & \textbf{0.280} & 0.514 & 0.32  & Bad photometry \\
    527   & 0.385 &   -    & 0.32  & Unconvincing \\
    537   & 0.353 & \textbf{0.287} & 0.30  &  \\
    865   & 0.278 & \textbf{0.234} & 0.24  &  \\
    1216  & \textbf{0.215} &   -    & 0.24  & RedMaPPer incompleteness\\
        \hline 

    \end{tabular}%
    }
    Note. The final correct redshift marked by \cite{rozo14} is written in bold.
  
\end{table}

\subsection{Validation Using Confirmed Planck Clusters}
\label{sec:ps-method-validation}

In \Fref{fig:ps-photoz-depth} we plot the redshift lower limit \zlimit\ versus the measured redshift of the candidates (using spectroscopic redshifts for those previously confirmed clusters).  We mark the successful validation clusters in blue, the validation clusters for which the redshift measurement failed in red, and the new candidates in green.   The dashed red line indicates where the \zlimit\ is equal to the cluster redshift.  Candidates that lie below this line have \ac{ps1} data that are sufficiently deep given the actual cluster redshift that we expect to extract a robust photo-z.  Candidates above the line would benefit from deeper imaging data, and for this reason we flag them as ``shallow''.

Beyond the redshift limit, we can reliably assign a redshift for some candidates, and this is not surprising.  The model we adopt in estimating the redshift lower limit \zlimit\ assumes a particular cluster mass, and many \planck\ candidates are indeed even more massive.  Also, our model does not account for the scatter in the expected number of red galaxies in a cluster at a particular redshift and mass.  In general, we would expect the photo-z's for these systems to be less robust, and indeed, we find that these systems show larger photometric redshift errors than the rest of the candidates.  

\begin{figure}
\includegraphics[width=3.4in]{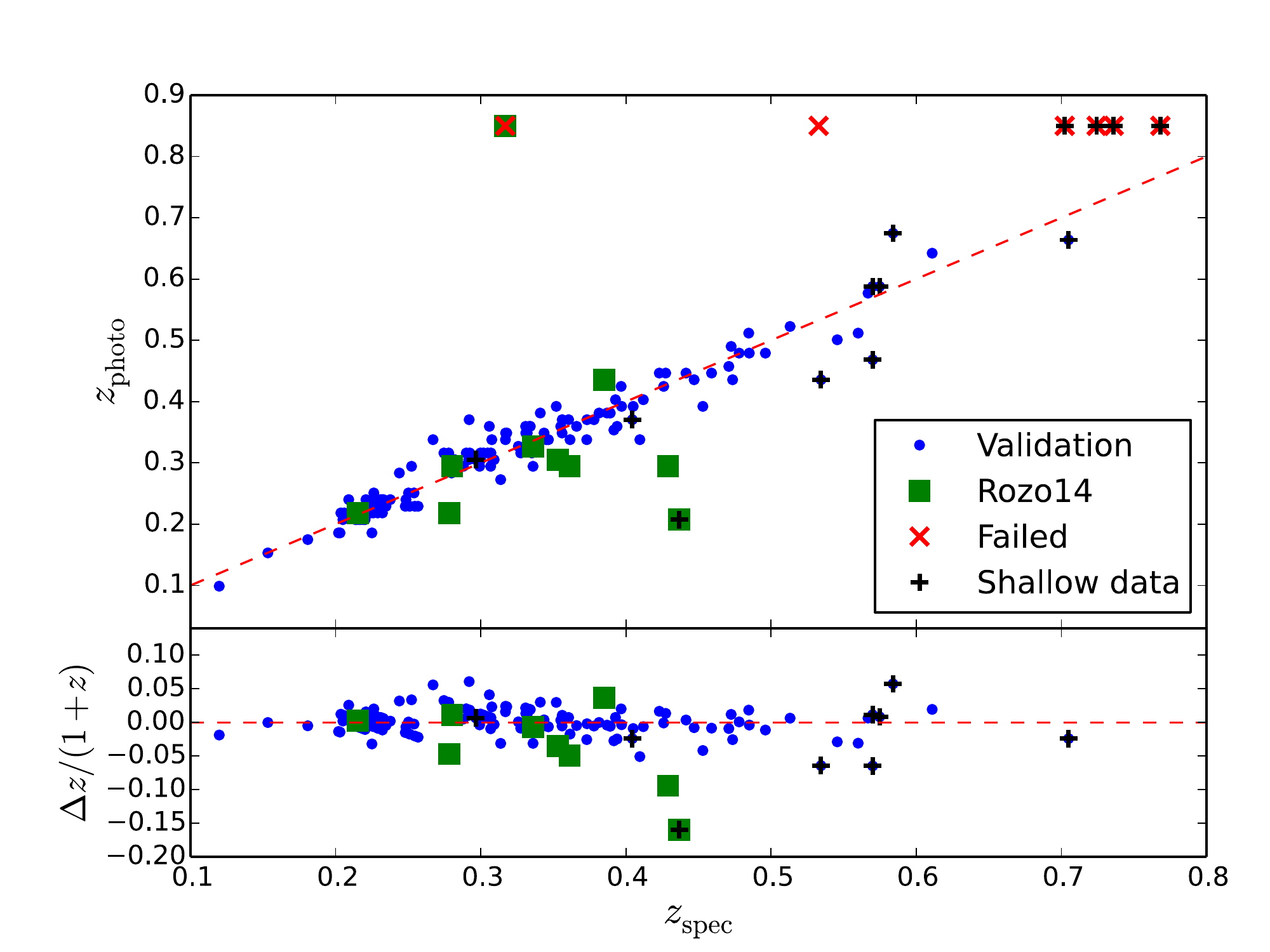}
  \caption[Photo-z validation.]{The \ac{photoz} measurements for \planck\ confirmed clusters plotted versus the spectroscopic redshifts (blue points).  The red crosses mark the failures in our photo-z estimation.  The black crosses mark the clusters whose redshifts are higher then the redshift limits, and the green squares marks the outliers examined in \cite{rozo14}.}
  \label{fig:ps-photoz-valid}
\end{figure}

The blinded photo-z estimation method fails to recover 6 of the 150 \planck\ confirmed clusters in the validation sample. Four of these cases correspond to clusters with redshifts above 0.7, which
are beyond the redshift lower limits \zlimit\ estimated from the depths of the \ac{ps1} data.  The other two failures are at redshifts below the estimated redshift lower limit.  One of these is \planck\ 484, which is a low-z cluster which is physically offset from the \ac{sze} detection by more than 5 arcmin.  In this case we repeat the analysis after recentering on the correct position and recover the \planck\ redshift.  The last failure corresponds to the cluster \planck\ 556 which is at a redshift of $z\approx 0.71$.  We note that in this case there is a low significance detection in the likelihood histogram, but we were not able to confirm it as a cluster.  A possible explanation is that this system has a somewhat lower mass than the characteristic mass we adopt in estimating the redshift lower limit.  Indeed, we find that both of these failed systems have relatively low values of $\ysz$, suggesting that they are lower mass systems.  Given the overall success (148/150) of the validation set, we are satisfied that if our depth estimate indicates we should be able to measure a cluster photometric redshift we will be able to do that with good reliability.

We also note that \cite{rozo14} present a comparison of the \planck\ redshifts with the RedMaPPer result based on \ac{sdss} data.  We cross-match the validation sample used here with the 3\,$\sigma$ outliers from table~1 of \cite{rozo14} and present the result in \Fref{tab:ps-cmp-rozo}.  Our results for the outliers are generally more consistent with the results from \cite{rozo14}.  

After estimating the redshifts for all candidates, we compare the photometric redshifts of the validation clusters with their spectroscopic redshifts and present this distribution in \Fref{fig:ps-photoz-valid}. After removing the failures and the questionable clusters identified in \cite{rozo14}, we are left with 135 \planck\ clusters.  We measure the \ac{rms} scatter defined as $(z_\text{photo}-z_\text{spec}) /(1+z_\text{spec})$ using the full spectroscopic cluster sample to be $0.023$.  We note that the redshift error distribution has a slight bias (0.003) that can be characterized empirically by a linear model.  We apply the bias correction to the measured candidate redshift values when quoting the final photo-z estimation.  After applying this bias correction, we obtain an \ac{rms} value of 0.022.  This value compares favorably with that of \cite{song12b} who measure an \ac{rms} scatter for three different photometric redshift estimation methods of between 0.028 and 0.024. We are satisfied that the measured \ac{rms} in this work demonstrates our ability to measure photometric redshifts for the \planck\ cluster candidates with the \ac{ps1} data.

Similar to \cite{song12b}, we estimate the final \ac{photoz} uncertainty as the quadrature sum of the measurement uncertainty and an intrinsic or systematic uncertainty $\delta_\text{sys}$: $\Delta^2 z_\text{phot}=\delta^2_{z\text{phot}}+\delta^2_\text{sys}$.  We find $\delta_\text{sys}=0.007$ by requiring that the reduced $\chi^{2}=1$ of the photometric redshifts about the spectroscopic redshifts for the validation ensemble.

\subsection{Results from Random Sky Regions}
Random superposition is one source of contamination in our analysis.  Given the large search radius (5~arcmin), the chance to associate an SZE selected candidate to a lower mass optical system is higher than the in our previous experience with the \ac{spt} sample.  Thus, we test our confirmation procedure against randomly selected sky regions to estimate the contamination rate.

We select 60 random candidate positions lying within an large equatorial region we are processing for other purposes; we produce coadds and calibrated catalogs in the same manner as for the real \planck\ candidates.  Then we search for optical counterparts around all random positions, and-- where possible-- estimate redshifts.  Out of the 60 random positions we identify six candidates that exhibit weak significance in their likelihood distributions and pass our detection threshold.  Further, two of these pass the second round visual examination where we require a clustered collection of galaxies.  Using SIMBAD, we find that one of them is a known cluster identified by \cite{propris2002} in the 2dF survey, but the other candidate is not associated with any previously known cluster.  The results of this test indicate that our method applied to \planck\ candidates and \ac{ps1} data suffers from a contamination rate of approximately $\sim$3~per cent.

\subsection{Results from the Planck Candidates Sample}
We are able to identify optical counterparts and measure redshifts for 60 of the full sample of 237 \planck\ candidates.  The \planck\ ID, the \ac{bcg} sky position ($\alpha_\text{BCG}$, $\delta_\text{BCG}$), the photometric redshift measurement and the redshift lower limit \zlimit\ are presented in \Fref{tab:ps-confirmed}.  An additional 83 candidates are located so close to the Galactic plane (see \Fref{fig:ps-whole-sky}) that we can not reliably assign a redshift or a redshift lower limit due to the high stellar density.  For the remaining 94 candidates, we are unable to identify an optical counterpart and we provide only redshift lower limits \zlimit\ that reflect the depths of the catalogue at those candidate locations.  This information together with the \planck\ ID is presented for each candidate in \Fref{tab:ps-unconfirmed}.

Nineteen of the confirmed candidates are in \planck\ Class~1 (c.f. \Fref{sec:ps-data-planck} for the \planck\ classification), whereas there are only three Class~1 candidates remaining in the 94 candidates.  This shows that our algorithm has confirmed most of the reliable detections from the \planck\ catalogue.  And the three remaining candidates may reside at redshifts beyond our redshift limits where deeper imaging is needed.

Using contamination estimates from the \planck\ collaboration \citep{planck13-29} together with the number of total \planck\ candidates and previously confirmed clusters, we estimate that only 45~per cent of our sample ($\sim$110) should be real clusters.  If we take our confirmed sample of 60 clusters together with 45~per cent of the 83 candidates lying in fields with high stellar contamination, we have accounted for 98 of our estimates 110 expected real clusters.  Thus, these numbers suggest that as many as 12 of our 94 unconfirmed candidates would likely turn out to be real clusters lying at redshifts beyond the redshift lower limits \zlimit\ we present.

Note that because the contamination rate is higher in the \planck\ catalogue in regions of high Galactic dust \cite[]{planck13-29}, the number of potentially unconfirmed clusters in the 83 candidates close to the Galactic plane may be less than our estimate.  This introduces additional uncertainty into our estimate of the expected number of unconfirmed candidates lying at $z>$\zlimit.  

Recently, \cite{planck14-26} published newly confirmed clusters using data from the Russian-Turkish~1.5~m telescope and 6~m Bolshoy Telescope Azimutal'ny of the Special Astrophysical Observatory of the Russian Academy of Sciences.  They confirmed 41 new \planck\ \ac{sze} clusters.  We cross match our sample with their results, finding that 11 clusters are in a good agreement and two others (candidates 383 and 618) exhibit large discrepancies ($\Delta z>0.1$).  In both of these cases our results prefer lower redshifts.  For the remaining 28 confirmed systems, we mark 16 as lying in star fields, and the rest are not fully covered in the \ac{ps1} data.

\section{Conclusions}\label{sec:ps-conclusion}
We study 237 unconfirmed \planck\ cluster candidates that overlap the \ac{ps1} footprint.  We describe the production of science ready catalogues and present the distribution of measured depths and photometric quality for this ensemble of cluster candidates.  We summarise our method for estimating cluster photometric redshifts and describe a method for estimating a redshift lower limit \zlimit\ beyond which we would not expect to be able to have confirmed the cluster in the \ac{ps1} data.  This method uses what we know about \ac{sze} selected massive clusters from SPT together with the measured depths of the \ac{ps1} catalogues.

We validate our photometric redshift estimation with a sample of 150 \planck\ confirmed clusters.  In this test, we fail to detect four  clusters that are beyond the redshift limit of the \ac{ps1} data, and two clusters that are within the redshift limits given the \ac{ps1} data quality.  We find  that 6 out of 10 previously identified clusters exhibiting large redshift discrepancies when comparing the \planck\ and \citet{rozo14} results exhibit redshifts that are more consistent with the \citet{rozo14} result.  For the remaining clusters, we achieve an overall redshift scatter of $(z_\text{photo}-z_\text{spec}) /(1+z_\text{spec})\sim0.022$.  We also examine the false detection rate due to random superposition of low mass galaxy systems.   Using 60 random sky regions, we find a contamination rate of $\sim$ 3 per cent, indicating that this fraction of our confirmed sample may be contaminated.

Using these data products and methods, we measure photometric redshifts for 60 \planck\ candidates.  The newly confirmed clusters span a redshift range $0.06\lt z\lt0.69$ with a median redshift $z_\text{med}=0.31$, which is consistent with the redshift distribution presented for the previously confirmed sample of \planck\ selected clusters.  This sample of 60 newly confirmed clusters increases the total number of new, \planck\ discovered clusters from 178 to 238, bringing the total \planck\ cluster sample -- including those discovered in previous surveys -- to 921 \citep{planck13-29}.

We exclude 83 of the remaining candidates because of high stellar contamination due to their position close to the Galactic plane.  For these systems we cannot obtain reliable photometric redshifts or estimate redshift lower limits with the current data.  We are unable to find optical counterparts or estimate photometric redshifts for the last 94 candidates in our sample.  For each of these we present a redshift lower limit \zlimit, but the majority of these systems are expected to be noise fluctuations.  

Using contamination estimates from the \planck\ collaboration \citep{planck13-29} we estimate that $\sim$12 of the 94 unconfirmed candidates could turn out to be real clusters lying at redshifts beyond the redshift lower limits \zlimit\ we present.  Confirming these systems will require short exposures on 4-m or 6.5-m class telescopes.  Additional \planck\ candidates can be obtained by mining the newly available \ac{des} data in the southern celestial hemisphere.  The \ac{des} depths are adequate to identify the optical counterparts and measure redshifts for high mass clusters out to $z\sim1.2$ (Hennig et al, in preparation).%

\section*{Acknowledgments}

The Munich group at LMU is supported by the DFG through TR33 ``The Dark Universe'' and the Cluster of Excellence ``Origin and Structure of the Universe''.  
The data processing has been carried out on the computing facilities of the Computational Center for Particle and Astrophysics (C2PAP), which is supported by the Cluster of Excellence. We want to thank H.\ H.\ Head from Austin Peay state university, with whom we initiated this project. Also we would like to thank J.\ Dietrich and D.\ C.\ Gangkofner at LMU for helpful discussions. 

The Pan-STARRS1 Surveys (PS1) have been made possible through contributions of the Institute for Astronomy, the University of Hawaii, the Pan-STARRS Project Office, the Max-Planck Society and its participating institutes, the Max Planck Institute for Astronomy, Heidelberg and the Max Planck Institute for Extraterrestrial Physics, Garching, The Johns Hopkins University, Durham University, the University of Edinburgh, Queen's University Belfast, the Harvard-Smithsonian Center for Astrophysics, the Las Cumbres Observatory Global Telescope Network Incorporated, the National Central University of Taiwan, the Space Telescope Science Institute, the National Aeronautics and Space Administration under Grant No. NNX08AR22G issued through the Planetary Science Division of the NASA Science Mission Directorate, the National Science Foundation under Grant No. AST-1238877, the University of Maryland, and Eotvos Lorand University (ELTE).

\begin{table}
  \centering
 \caption{Sky positions and redshifts of \planck\ candidates.}
 \label{tab:ps-confirmed}
    \begin{tabular}{rrrrrr}
\hline
ID    & $z_\text{phot}$ &  $\Delta z_\text{phot}$ &$\alpha_\text{BCG}$  & $\delta_\text{BCG}$ & \zlimit \\ \hline
43    & 0.077 & 0.007 & 253.0509 & $-0.3377$ & 0.58 \\
59    & 0.284 & 0.013 & 313.5165 & $-22.8076$ & 0.59 \\
66    & 0.533 & 0.250 & 330.7982 & $-24.6406$ & 0.55 \\
70    & 0.284 & 0.020 & 257.9357 & 7.2559 & 0.62 \\
83    & 0.425 & 0.022 & 344.8704 & $-25.1154$ & 0.57 \\
111   & 0.251 & 0.029 & 323.2163 & $-12.5426$ & 0.59 \\
116   & 0.479 & 0.037 & 266.7882 & 17.1839 & 0.58 \\
126   & 0.240 & 0.021 & 316.1941 & $-4.7623$ & 0.56 \\
133   & 0.229 & 0.034 & 273.5555 & 18.2843 & 0.57 \\
142   & 0.360 & 0.016 & 219.4179 & 30.2001 & 0.72 \\
142$^*$ & 0.170 & 0.010 & 219.4585 & 30.4253 & 0.72 \\
143   & 0.240 & 0.007 & 252.5850 & 26.9726 & 0.70 \\
149   & 0.544 & 0.070 & 335.0728 & $-12.1916$ & 0.58 \\
150   & 0.381 & 0.031 & 347.4625 & $-18.3324$ & 0.57 \\
157   & 0.218 & 0.046 & 359.2370 & $-22.7796$ & 0.56 \\
209   & 0.403 & 0.035 & 313.2155 & 17.9064 & 0.48 \\
212   & 0.403 & 0.007 & 257.6559 & 40.4314 & 0.66 \\
213   & 0.686 & 0.133 & 229.0082 & 39.7408 & 0.69 \\
218   & 0.273 & 0.034 & 319.8591 & 15.3518 & 0.54 \\
257   & 0.436 & 0.054 & 242.2561 & 50.0867 & 0.68 \\
261   & 0.088 & 0.007 & 290.8001 & 48.2705 & 0.57 \\
262   & 0.479 & 0.022 & 3.8511 & $-17.5108$ & 0.64 \\
282   & 0.316 & 0.021 & 324.4442 & 35.5975 & 0.40 \\
289   & 0.099 & 0.007 & 300.8065 & 51.3474 & 0.49 \\
305   & 0.207 & 0.012 & 352.1669 & 7.5801 & 0.61 \\
314   & 0.262 & 0.026 & 257.4693 & 62.3689 & 0.64 \\
375   & 0.099 & 0.034 & 283.0395 & 72.9927 & 0.64 \\
383   & 0.360 & 0.028 & 284.2933 & 74.9421 & 0.64 \\
420   & 0.229 & 0.018 & 0.3115 & 50.2756 & 0.45 \\
509   & 0.284 & 0.031 & 140.0173 & 70.8205 & 0.60 \\
522   & 0.077 & 0.007 & 27.8319 & 10.8141 & 0.64 \\
529   & 0.110 & 0.007 & 99.4772 & 66.8518 & 0.51 \\
543   & 0.131 & 0.046 & 129.9560 & 62.4101 & 0.50 \\
553   & 0.349 & 0.031 & 100.1444 & 57.7460 & 0.54 \\
554   & 0.305 & 0.019 & 36.2339 & 8.8299 & 0.60 \\
554$^*$ & 0.310 & 0.019 & 36.1653 & 8.8983 & 0.60 \\
575   & 0.294 & 0.088 & 119.3808 & 52.6829 & 0.60 \\
576   & 0.153 & 0.014 & 150.4115 & 50.0149 & 0.65 \\
612   & 0.349 & 0.034 & 60.7362 & 9.7414 & 0.63 \\
618   & 0.370 & 0.365 & 100.7427 & 31.7503 & 0.48 \\
679   & 0.251 & 0.018 & 48.8412 & $-18.2062$ & 0.57 \\
682   & 0.381 & 0.036 & 112.5014 & 11.9483 & 0.63 \\
699   & 0.381 & 0.086 & 146.1786 & 19.4666 & 0.50 \\
701   & 0.316 & 0.045 & 179.8416 & 26.4511 & 0.66 \\
723   & 0.327 & 0.027 & 117.2153 & 1.1111 & 0.62 \\
725   & 0.305 & 0.028 & 32.2630 & $-27.5107$ & 0.58 \\
735   & 0.131 & 0.049 & 78.7192 & $-19.9555$ & 0.61 \\
736   & 0.664 & 0.007 & 48.7537 & $-27.3029$ & 0.63 \\
743   & 0.381 & 0.066 & 160.2901 & 17.5098 & 0.61 \\
748   & 0.099 & 0.007 & 112.8076 & $-7.8093$ & 0.68 \\
752   & 0.294 & 0.013 & 120.4230 & $-4.0614$ & 0.50 \\
778   & 0.403 & 0.028 & 94.7096 & $-23.5784$ & 0.57 \\
828   & 0.251 & 0.044 & 126.6873 & $-23.2611$ & 0.53 \\
837   & 0.436 & 0.307 & 131.7742 & $-21.9784$ & 0.56 \\
860   & 0.338 & 0.020 & 142.9920 & $-20.6231$ & 0.56 \\
913   & 0.392 & 0.067 & 158.8869 & $-20.8495$ & 0.57 \\
978   & 0.327 & 0.036 & 175.3720 & $-21.6974$ & 0.66 \\
1001  & 0.349 & 0.025 & 178.5667 & $-26.1542$ & 0.57 \\
1080  & 0.207 & 0.137 & 195.4422 & $-12.0830$ & 0.58 \\
1159  & 0.055 & 0.007 & 201.6415 & 11.3018 & 0.64 \\
1178  & 0.294 & 0.028 & 223.1756 & $-18.5844$ & 0.67 \\
1189  & 0.305 & 0.021 & 216.3013 & $-4.9427$ & 0.60 \\
1189$^{*}$  & 0.330 & 0.021 & 216.3943 & $-5.0097$ & 0.68 \\ \hline
\multicolumn{6}{l}{Note. Multiple detections are marked with $^{*}$.}
\end{tabular}
\end{table}

\begin{table*} 
\caption[Unconfirmed Planck candidate.]{Unconfirmed \planck\ cluster candidates with redshift lower limits \zlimit.}\label{tab:ps-unconfirmed}
  \begin{center} 
\begin{tabular}{cccccccccccc} \hline 
\planck\ ID & \zlimit & \planck\ ID & \zlimit & \planck\ ID & \zlimit & \planck\ ID & \zlimit & \planck\ ID & \zlimit & \planck\ ID & \zlimit \\ \hline 
38 & $0.66$ & 58 & $0.58$ & 84 & $0.62$ & 86 & $0.67$ & 90 & $0.67$ & 104 & $0.58$ \\
176 & $0.56$ & 193 & $0.59$ & 211 & $0.69$ & 251 & $0.62$ & 271 & $0.64$ & 279 & $0.70$ \\
298 & $0.57$ & 306 & $0.65$ & 310 & $0.56$ & 311 & $0.53$ & 317 & $0.44$ & 318 & $0.56$ \\
320 & $0.52$ & 331 & $0.55$ & 346 & $0.73$ & 361 & $0.66$ & 370 & $0.48$ & 372 & $0.60$ \\
373 & $0.57$ & 377 & $0.67$ & 381 & $0.61$ & 382 & $0.52$ & 387 & $0.53$ & 395 & $0.57$ \\
397 & $0.53$ & 398 & $0.55$ & 412 & $0.64$ & 424 & $0.55$ & 425 & $0.65$ & 437 & $0.57$ \\
458 & $0.65$ & 476 & $0.58$ & 490 & $0.67$ & 497 & $0.62$ & 504 & $0.70$ & 507 & $0.50$ \\
517 & $0.68$ & 534 & $0.55$ & 538 & $0.72$ & 544 & $0.60$ & 549 & $0.50$ & 555 & $0.61$ \\
564 & $0.58$ & 566 & $0.58$ & 580 & $0.49$ & 586 & $0.55$ & 597 & $0.49$ & 605 & $0.52$ \\
611 & $0.41$ & 616 & $0.67$ & 624 & $0.57$ & 625 & $0.51$ & 626 & $0.54$ & 629 & $0.74$ \\
651 & $0.59$ & 652 & $0.59$ & 658 & $0.67$ & 663 & $0.62$ & 684 & $0.51$ & 695 & $0.58$ \\
712 & $0.66$ & 722 & $0.57$ & 755 & $0.51$ & 766 & $0.48$ & 775 & $0.52$ & 791 & $0.66$ \\
792 & $0.69$ & 798 & $0.60$ & 809 & $0.65$ & 845 & $0.58$ & 864 & $0.63$ & 884 & $0.65$ \\
886 & $0.58$ & 900 & $0.57$ & 909 & $0.63$ & 928 & $0.69$ & 992 & $0.66$ & 1070 & $0.66$ \\
1122 & $0.66$ & 1132 & $0.63$ & 1152 & $0.65$ & 1171 & $0.50$ & 1175 & $0.49$ & 1198 & $0.68$ \\
1199 & $0.67$ & 1212 & $0.68$ & 1217 & $0.68$ & 1221 & $0.64$ \\ \hline
\end{tabular}\end{center}
\end{table*}

\bibliographystyle{mn2e}
\bibliography{ref}
\end{document}